\documentclass[aps,prl,twocolumn,superscriptaddress,showpacs]{revtex4}

\usepackage{graphicx}
\usepackage{float}
\usepackage{amsfonts}
\usepackage{amsmath}
\usepackage{amssymb}
\usepackage{enumerate}
\usepackage{color}
\usepackage{mciteplus}
\usepackage{natbib}

\usepackage{epstopdf} 

\newcommand{\ket}[1]{\left\vert{#1}\right\rangle}
\newcommand{\be}{\begin{equation}}
\newcommand{\ee}{\end{equation}}

\newcommand{\ba}{\begin{array}}
\newcommand{\ea}{\end{array}}
\newcommand{\bqa}{\begin{eqnarray}}
\newcommand{\eqa}{\end{eqnarray}}

\begin{document}

\title{Observation of quantum interference in the plasmonic Hong-Ou-Mandel effect}


\author{G. Di Martino}
\affiliation{Experimental Solid State Group, Imperial College London, Blackett Laboratory, SW7 2AZ London, UK}
\author{Y. Sonnefraud}
\affiliation{Experimental Solid State Group, Imperial College London, Blackett Laboratory, SW7 2AZ London, UK}
\author{M. S. Tame}
\email{markstame@gmail.com}
\affiliation{Experimental Solid State Group, Imperial College London, Blackett Laboratory, SW7 2AZ London, UK}
\affiliation{University of KwaZulu-Natal, School of Chemistry and Physics, 4001 Durban, South Africa}
\author{S. K\'ena-Cohen}
\affiliation{Experimental Solid State Group, Imperial College London, Blackett Laboratory, SW7 2AZ London, UK}
\author{F. Dieleman}
\affiliation{Experimental Solid State Group, Imperial College London, Blackett Laboratory, SW7 2AZ London, UK}
\author{\c{S}. K. \"Ozdemir}
\affiliation{Department\,of\,Electrical\,and\,Systems\,Engineering, Washington University, St. Louis, MO 63130, USA}
\author{M. S. Kim}
\affiliation{Quantum Optics and Laser Science Group, Imperial College London, Blackett Laboratory, SW7 2AZ London, UK}
\author{S. A. Maier}
\email{s.maier@imperial.ac.uk}
\affiliation{Experimental Solid State Group, Imperial College London, Blackett Laboratory, SW7 2AZ London, UK}

\begin{abstract}
We report direct evidence of the bosonic nature of surface plasmon polaritons (SPPs) in a scattering-based beamsplitter. A parametric down-conversion source is used to produce two indistinguishable photons, each of which is converted into a SPP on a metal-stripe waveguide and then made to interact through a semi-transparent Bragg mirror. In this plasmonic analog of the Hong-Ou-Mandel experiment, we measure a coincidence dip with a visibility of 72\%, a key signature that SPPs are bosons and that quantum interference is clearly involved.
\end{abstract}


\maketitle


Nanophotonic systems based on plasmonic components are currently attracting considerable attention due to the novel ways in which the electromagnetic field can be localised and controlled~\cite{Gramotnev2010,Giannini2010}. In the classical regime, a wide range of applications are being pursued, including nano-imaging~\cite{Kawata2009}, biosensing~\cite{Anker2008} and solar cells~\cite{Atwater2010}. Recently, researchers have started to investigate plasmonics in the quantum regime~\cite{Tame2013}. Devices have been proposed for a variety of applications in quantum information science~\cite{Akimov2007,Chang2007,deLeon2012,Tame2013}. Despite the remarkable progress made so far, there are many fundamental aspects of quantum plasmonic systems that remain unexplored. One key property is the bosonic character of single surface plasmon polaritons (SPPs). The quasi-particle nature of SPPs, consisting of a photon coupled to a density wave of electrons, makes them an unusual type of excitation. While it is generally accepted that in theory SPPs are bosons, as of yet, the quantum statistical behaviour of SPPs has not been clearly demonstrated experimentally. The bosonic nature of \emph{photons} was explicitly verified in the seminal experiment of Hong, Ou and Mandel~\cite{Hong1987}. Recent work using plasmonic waveguides has hinted that SPPs are bosons by observing the preservation of properties of the photons used to excite them~\cite{Altewischer2002, Fasel2005, Fasel2006, Huck2009}, and the Hong-Ou-Mandel (HOM) effect, both indirectly using a photonic beamsplitter~\cite{Fujii2012} and directly using a plasmonic beamsplitter~\cite{Heeres2013}. However, the question as to whether quantum interference is involved remains open due to the low HOM interference observed, which can be obtained via classical interference of light~\cite{Rarity2005,Ghosh1987,Mandel1999,Ballester2010,Kim2013}. In order to verify the bosonic nature of single excitations in the quantum regime it is vital to observe quantum interference~\cite{Rarity2005,Ghosh1987,Mandel1999}. 

In this work we report the first observation of quantum interference in the HOM effect for SPPs. We have used spontaneous parametric down-conversion (SPDC) to produce two indistinguishable photons~\cite{Burnham1970,Hong1986}, each of which is converted into a SPP on separate metal-stripe waveguides~\cite{Lamprecht2001,Zia2005,DiMartino2012}. This approach alleviates difficulties related to indistinguishability when using quantum emitters as single-SPP sources~\cite{Kolesov2009,Cuche2010,Mollet2012}. The generated SPPs interact on a beamsplitter via a scattering process~\cite{Ditlbacher2002,Weeber2005,Gonzalez2006} and we find that the they exhibit the distinct bunching effect expected for bosons, with the results clearly showing quantum interference is involved. 


\begin{figure*}[t]
\centering  
\includegraphics[width=12.7cm]{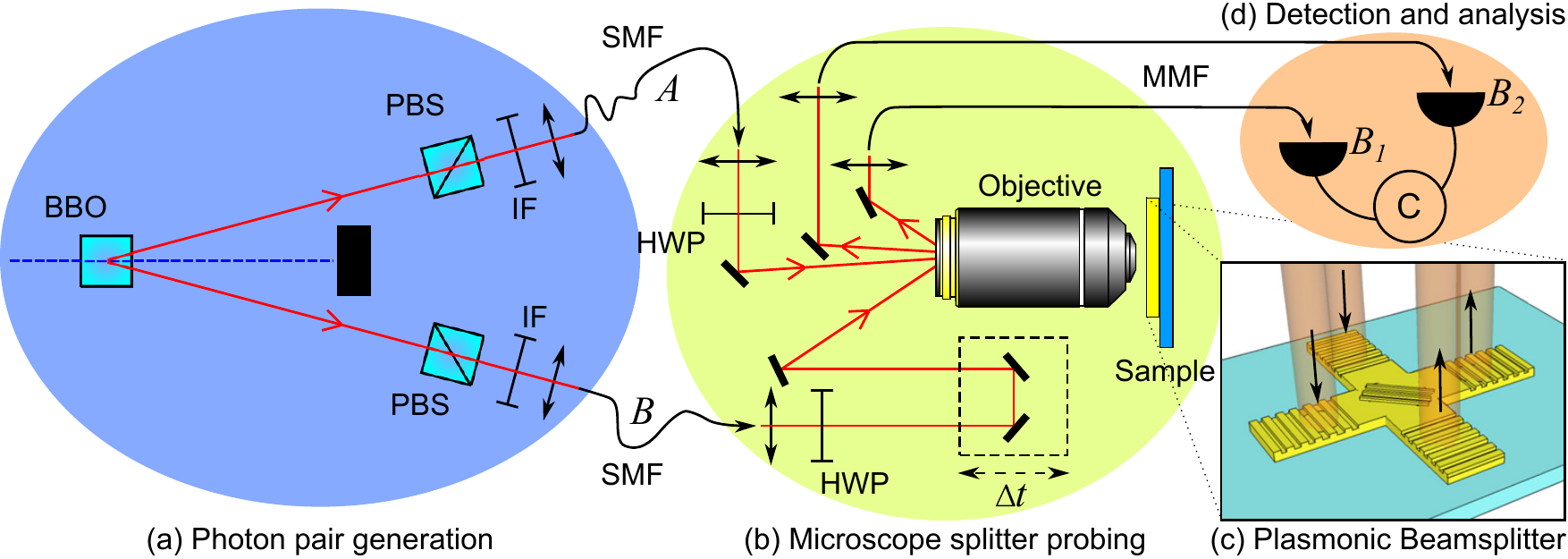}
\caption{Experimental setup. (a) Photon pair generation. Photon pairs are generated via spontaneous parametric down-conversion using a pump laser focused onto a Beta Barium Borate (BBO) crystal and filtered using interference filters (IF). Each photon is coupled into a single-mode fiber (SMF). (b) Microscopy. The photons from the SMFs are collimated and half-wave plates (HWPs) are used to optimize surface plasmon polariton (SPP) excitation. A time delay is introduced on one path. (c) Plasmonic beamsplitter. The photons are focused onto separate spots on the input gratings using a microscope objective. The beams at the output gratings are collected and coupled into multi-mode fibers (MMFs). (d) Detection and analysis. The outputs of the MMFs are sent to avalanche photodiodes, $B_1$ and $B_2$, where coincident detection events are measured.}
\label{expsetup}
\end{figure*}

\begin{figure}[t]
\centering  
\includegraphics[width=6.3cm]{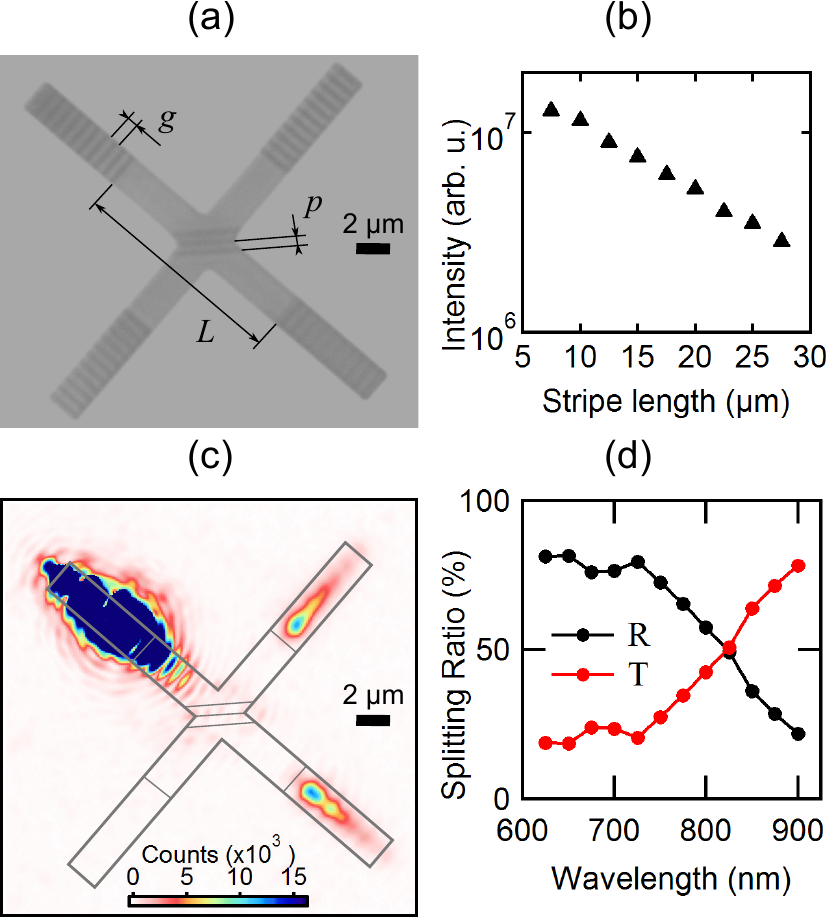}
\caption{Plasmonic beamsplitter. (a) Optical image of the beamsplitter. The in/out gratings consist of eleven ridges, each being repeated at an increment of $g=$~620~nm from the waveguide end. The distance between gratings is $L=12.5~\mu$m. The Bragg reflector is made out of three ridges with a center-to-center distance of $p=$~500~nm. (b) Intensity out-coupled from a single waveguide as a function of length when excited by a laser at 808~nm using the method in ref.~\cite{DiMartino2012}. (c) Optical image of the splitter when the SPPs are excited by a laser at 808~nm focused on the top-left grating. Light is out-coupled at the top-right and bottom-right with almost identical intensity. The integration time has been adjusted to give a reasonable contrast for the output, leading to a saturation at the input caused by the scattered field. (d) Transmission $T$ and reflection $R$ as a function of wavelength.} 
\label{FigSplitter}
\end{figure}

{\it Experimental setup.--} The setup used to conduct our investigation is shown in Fig.~\ref{expsetup}~(a). Here, photon pairs are generated at a wavelength of 808~nm using a 100~mW continuous wave laser ($\lambda=404$~nm) focused onto a Beta Barium Borate (BBO) crystal cut for type-I SPDC~\cite{Burnham1970}. Phase matching conditions lead to the photons from a given pair being emitted into antipodal points of a cone with an opening angle of 6$^\circ$~\cite{Hong1986}. Fig.~\ref{expsetup}~(a) shows the antipodal points chosen are in-plane. Polarizing beam splitters (PBSs) in the paths of the down-converted beams remove any parasitic light with the incorrect polarization. Interference filters (IFs) with a central wavelength of 800~nm and 22~nm bandwidth are placed in both paths to spectrally select out the down-converted photons. The photons are injected into single-mode fibers (SMFs).  After collimation of the output from the fibers, the polarization is adjusted using half-wave plates (HWPs) to maximize the excitation of SPPs on the sample. The polarization dependence of the SPP excitation efficiency is the same as in ref.~\cite{DiMartino2012}. In order to control how well the SPPs generated from the photons interfere with each other we introduce a degree of distinguishability. Their spatial and spectral characteristics are closely matched by the SMFs and IFs, so a time delay is introduced in one path using a motorized delay line of distance $d$. This provides a variable delay of $\Delta t = d/c$ between the single-photon wavepackets, so that the arrival time of the SPPs at the plasmonic beamsplitter can be controlled and a degree of distinguishability introduced. The photons are focused onto separate gratings (spot size 2 $\mu$m) at the inputs of an X-shaped plasmonic beamsplitter, shown in Fig.~\ref{expsetup}~(c), by a microscope objective (100$\times$, NA~0.8) and converted into SPPs due to phase-matching conditions~\cite{DiMartino2012,Tame2008}. The SPPs propagate along the waveguides, passing through the central body of the beamsplitter where they interfere via a scattering process. After scattering, they reach the output gratings and are converted back into light. Fig.~\ref{FigSplitter}~(c) shows the SPP intensity from the two output gratings. Multi-mode fibers (MMFs) collect this out-coupled light, directing it to silicon avalanche photodiode (APD) detectors $B_1$ and $B_2$, which monitor the arrival of the photons. Detection events are time-tagged (PicoQuant Hydraharp 400), coincidences are evaluated within a $t_c = 2$~ns time window and reported error bars correspond to standard deviations.


{\it Beamsplitter characterization.--} The plasmonic beamsplitter consists of two 2~$\mu$m wide, 70~nm thick gold stripe waveguides that cross at a right angle at their midpoint, as shown in Fig.~\ref{FigSplitter}~(a). These waveguides support a single low-loss leaky SPP mode~\cite{Zia2005} and a number of short-range bound modes~\cite{Lamprecht2001}. The beamsplitter structure was defined on a glass substrate by electron beam lithography (EBL). A second EBL step is used to overlay 90~nm thick input/output gratings and central scattering elements, as described in ref.~\cite{KenaCohen2013}. The grating periodicity $g=620$~nm is chosen to couple effectively to the low-loss SPP mode. The SPP propagation length (the length at which the intensity decreases to 1/e of its original value) is $l$ = 12.4~$\pm$~0.3~$\mu$m. This has been measured on gold stripe waveguides of increasing length, as described in ref.~\cite{DiMartino2012} and shown in Fig.~\ref{FigSplitter}~(b). The distance between in-coupling and out-coupling gratings is $L=12.5~\mu$m. 

In the beamsplitter the splitting operation is obtained via a scattering process, in direct contrast to previous studies using coupled waveguides~\cite{Heeres2013}. The scattering element is a semi-transparent Bragg reflector, consisting of three ridges spaced by a distance $p=$~500~nm, deposited on the central part of the beamsplitter, as shown in Fig.~\ref{FigSplitter}~(a). Bragg reflectors such as the one we use have been studied extensively in the literature, mostly as effective mirrors in the one-dimensional case of normal incidence, both on an infinite interface and on plasmonic waveguides~\cite{Weeber2005}. Some reflectors have also been studied in the two-dimensional case with different structures, such as a grating made of nanoparticles~\cite{Ditlbacher2002,Quinten2008}, or with ridges on an infinite interface~\cite{Gonzalez2006}. We have chosen this Bragg reflector approach over coupled waveguides due to its compactness and the potential for multiple elements to be integrated: the zone over which the SPPs interact represents less than two wavelengths.

To obtain the wavelength dependence of the transmission and reflection, $T$ and $R$, of the Bragg element, we use light from a supercontinuum filtered to the appropriate wavelength and focused on one of the input gratings, {\it e.g.} the top-left grating, as shown in Fig.~\ref{FigSplitter}~(c). For each wavelength the intensity is integrated over the complete area of each output grating (top-right and bottom-right gratings in Fig.~\ref{FigSplitter}~(c)). The value of $T$ is the ratio between the intensity at the output grating directly opposite the input grating and the total intensity at both output gratings. While this method does not account for loss due to radiative scattering at the Bragg reflector or during SPP propagation, it gives the relative transmission $T$ of the beamsplitter. From the above, the relative reflection coefficient is then $R=1-T$. In order to maximize quantum interference in the HOM effect the beamsplitter must have $R=T=1/2$~\cite{Hong1987}. We have checked the splitting ratio for a range of wavelengths for optimal splitting. As shown in Fig.~\ref{FigSplitter}~(d), a Bragg reflector with ridges having a period of  500~nm gives $T=0.49 \pm 0.05$ for incident SPPs at $\lambda_0 = 808$~nm -- the wavelength of the photons used in our experiment. The value of $T$ remains the same for every input of the beamsplitter.

{\it Theoretical background.--} In the HOM experiment, due to phase matching conditions of the SPDC process, a single photon in path $i$ is well approximated by the quantum state $\ket{1}_i=\int d \omega_i \phi_i(\omega_i)\hat{a}_i^\dag(\omega_i)\ket{0}$,~\cite{Hong1986,Hong1987} where $\phi_i(\omega_i)$ is a normalized spectral amplitude of the photon, $\ket{0}$ is the vacuum state and $\hat{a}_i^\dag(\omega_i)$ is a creation operator, which together with the operator $\hat{a}_i(\omega_i)$ satisfies the bosonic commutation relation $[\hat{a}_i(\omega_i),\hat{a}_i^\dag(\omega_i')]=\delta(\omega_i-\omega_i')$~\cite{Loudon2000}. Taking the input state at the beamsplitter as $\ket{1}_A\ket{1}_B$ and applying the unitary transformations $\hat{a}_A^\dag(\omega)=i\sqrt{R}\hat{a}_{B_1}^\dag(\omega)+\sqrt{T}\hat{a}_{B_2}^\dag(\omega)$ and $\hat{a}_B^\dag(\omega)=\sqrt{T}\hat{a}_{B_1}^\dag(\omega)+i\sqrt{R} \hat{a}_{B_2}^\dag(\omega)$ leads to the output state 
\bqa
&&\eta_{in}\eta_{out}[i\sqrt{R}\sqrt{T}\ket{2}_{B_1}\ket{0}_{B_2}+i\sqrt{R}\sqrt{T} \ket{0}_{B_1}\ket{2}_{B_2} \nonumber \\ 
&&\qquad \qquad -R\ket{1}_{B_1}\ket{1}_{B_2}+T\ket{1}_{B_1}\ket{1}_{B_2}] \label{output}
\eqa
where $\eta_{in}$ ($\eta_{out}$) accounts for loss in the input (output) arms of the beamsplitter~\cite{Loudon2000}. In the ideal case, $R=T=1/2$ and the terms with one excitation in each output interfere destructively. This interference can only be seen in the quantum regime and leads to the output state
\be
\ket{1}_A\ket{1}_B \to \frac{1}{\sqrt{2}}(\ket{2}_{B_1}\ket{0}_{B_2}+\ket{0}_{B_1}\ket{2}_{B_2}). \label{HOM}
\ee
Thus, the photons display bosonic behaviour by bunching together. This bunching occurs regardless of the loss at the input and output stages, which only reduces the rate at which the process occurs. From Eq.~(\ref{HOM}), the probability of detecting a coincidence event where a photon is present at each output drops to zero when the photons interfere. On the other hand, when they are unable to interfere, {\it e.g.} due to their arrival time, each output state in Eq.~(\ref{output}) occurs with equal probability and the probability of detecting a coincidence is 1/2 (scaled by $(\eta_{in}\eta_{out})^2$). In the time domain, when $R=T=1/2$, Eq.~(\ref{output}) leads to a coincidence probability
$P(\Delta t) = (\eta_{in}\eta_{out})^2(1-{\text{sinc}}^2(\Delta t \cdot \Delta \omega/2))/2$.
Here, $\Delta t$ is the delay between the photons and top-hat amplitudes $\phi_i (\omega_i)$ are used with a FWHM of $\Delta \omega$. Thus, we have $P(0)=0$ and $P(\Delta t \gg \tau_c)=(\eta_{in}\eta_{out})^2/2$, where $\tau_c \sim 2 \pi/\Delta \omega$ is the photon coherence time and $\tau_c \ll t_c$, with $t_c$ the coincidence window of the detection events.


{\it Quantum interference.--} We first confirmed that the photons generated by our source exhibit the above-described HOM effect in a conventional beamsplitter. For this, we measured the output coincidences, as a function of time delay between the arrival of the input photons. At zero delay the coincidence rate drops to a minimum value, $N_{min}$, as expected. This drop is quantified using the visibility, $V_P=(N_{max}-N_{min})/N_{max}$~\cite{Ghosh1987,Rarity2005}, where $N_{max}$ is the maximum value of coincidences far from the dip center. We find $V_P = 0.67 \pm 0.05$. This value is limited by a number of factors, including the bandwidth of the IFs used in our experiment, the resolution of the time delay, the spatial mismatch between the modes of the photons at the beamsplitter and the deviation of the beamsplitter from the ideal case ($R=T=0.5$). Despite these factors, with a visibility larger than 0.5 we can confirm that the drop is due to \emph{quantum} interference~\cite{Rarity2005,Ghosh1987,Mandel1999}.

\begin{figure}[t]
\centering  
\includegraphics[width=6cm]{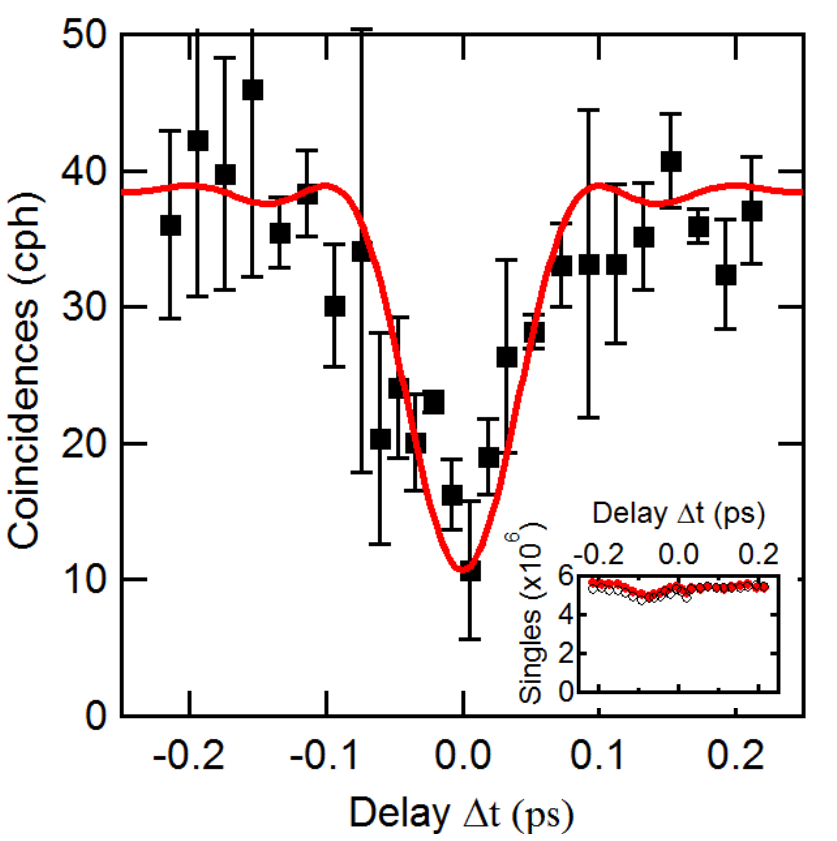}
\caption{Plasmonic Hong-Ou-Mandel dip. Black squares: coincidence rate as a function of time delay $\Delta t$. The red curve is a theoretical fit, $N(\Delta t)$, based on the coincidence probability $P(\Delta t)$ corrected for accidentals~\cite{supp}. From the theory fit we extract a coherence time for the SPPs of $\sim0.1$~ps, which is consistent with the coherence time obtained from the measured photonic dip. This confirms that during the photon-SPP conversion process the coherence properties of the single-photon wavepackets and the single SPPs are similar, and that the wavepacket has not been significantly altered by the conversion process or propagation. The inset shows the singles rates (in cph) as a function of $\Delta t$: red disks for detector $B_1$ and black circles for $B_2$. The visibility obtained from the plasmonic dip is $V_{SPP} = 0.72 \pm 0.07$.}
\label{PlasmonicDip}
\end{figure}

We then probed the plasmonic beamsplitter, as depicted in Fig.~\ref{expsetup}. When the coupling of single photons into the SPP waveguides is optimized, the count rate due to SPPs scattered by the output grating and detected by APD $B_{1,2}$ is $N_{B_{1,2}} \sim 5.5\times10^6$~cph, as shown in the inset of Fig.~\ref{PlasmonicDip}. The time-resolved correlation data shows an average number of coincidences of $54.6 \pm 1.1$ counts per hour (cph) far from zero time delay and $30.3 \pm 1.2$~cph at zero delay. A proportion of these counts are due to accidental coincidences from uncorrelated photon pairs which couple into the beamsplitter but do not correspond to true correlated pairs from the source, thus we subtract them from the overall counts~\cite{supp}. The average coincidence rate far from zero delay is then $37.7 \pm 1.0$~cph, as shown in Fig.~\ref{PlasmonicDip}. On the other hand, at zero delay we have $10.7 \pm 2.5$~cph, which leads to a visibility for the plasmonic HOM dip of $V_{SPP} = 0.72 \pm 0.07$. The observed dip confirms that single SPPs bunch together as bosons and as the visibility is larger than 0.5 this confirms quantum interference is involved in the bunching process~\cite{Rarity2005,Ghosh1987,Mandel1999}. The plasmonic visibility is again limited by a number of factors, including the bandwidth of the IF's and the time-delay resolution. 
The use of a narrower bandwidth is possible, giving photons with improved spectral definition for interference, however this comes at the expense of longer data collection times, where the data becomes sensitive to the coupling stability of the setup. The resolution of the time delay is limited by the accuracy of the translation stage, with shorter step sizes allowing for improved accuracy near the dip minimum. One reason for the improved visibility compared to the photonic case may be due to the integrated waveguide providing better spatial overlap of the modes at the beamsplitter~\cite{Politi2008}. Furthermore, any loss due to radiative scattering at the beamsplitter (measured as $<10$\%) occurs instantaneously and can be included within $\eta_{out}$, which does not play a role in reducing the visibility~\cite{Ballester2010}. Finally, note that when the coincidence rate drops as the single SPPs interfere, the count rate at each APD remains unchanged, as shown in Fig.~\ref{PlasmonicDip}. This is due to the small portion of \emph{pairs} of single excitations (compared to the total) that survive the process of propagation, splitting and out-coupling: in most cases at least one excitation from a pair will be lost. Therefore the count rate at each APD allows the efficiency of the system to be monitored to ensure the dip is not caused by loss fluctuations. 


{\it Summary.--} In this work we used single photons generated via parametric down conversion to excite single SPPs on a metallic stripe beamsplitter. The SPPs interacted via a scattering process and we directly observed the HOM effect. The SPPs showed a distinct bunching behaviour as expected for bosons, with the results clearly showing quantum interference is involved. Our investigation confirms the bosonic nature of single SPPs in the quantum regime and opens up new opportunities for controlling quantum states of light in ultra-compact nanophotonic plasmonic circuitry.

{\it Acknowledgments.--} This work was supported by the UK Engineering and Physical Sciences Research Council, the Leverhulme Trust, the European Office of Aerospace Research and Development (EOARD), and the Qatar National Research Fund (Grant NPRP 4-554-1-D84). We thank P. L. Knight, R. Oulton and A. Lupi for comments and discussions. SKO thanks L. Yang and F. Nori for support.


\end{document}